\pdfoutput=1
\RequirePackage{ifpdf}
\ifpdf 
\documentclass[pdftex]{sigma}
\else
\documentclass{sigma}
\fi

\usepackage{enumerate}

\begin{document}

\allowdisplaybreaks

\renewcommand{\thefootnote}{$\star$}

\renewcommand{\PaperNumber}{043}

\FirstPageHeading

\ShortArticleName{Functions characterizing the Ground State of the XXZ Spin-1/2 Chain}

\ArticleName{Functions Characterizing the Ground State of the\\ XXZ Spin-1/2 Chain in the Thermodynamic
Limit\footnote{This paper is a~contribution to the Special Issue in honor of Anatol Kirillov and Tetsuji Miwa.
The full collection is available at \href{http://www.emis.de/journals/SIGMA/InfiniteAnalysis2013.html}
{http://www.emis.de/journals/SIGMA/InfiniteAnalysis2013.html}}}

\Author{Maxime DUGAVE~$^\dag$, Frank G\"{O}HMANN~$^\dag$ and Karol Kajetan KOZLOWSKI~$^\ddag$}

\AuthorNameForHeading{M.~Dugave, F.~G\"{o}hmann and K.K.~Kozlowski}

\Address{$^\dag$~Fachbereich C~-- Physik, Bergische Universit\"at Wuppertal, 42097 Wuppertal, Germany}
\EmailD{\href{mailto:dugave@uni-wuppertal.de}{dugave@uni-wuppertal.de},
\href{mailto:goehmann@uni-wuppertal.de}{goehmann@uni-wuppertal.de}}

\Address{$^\ddag$~IMB, UMR 5584 du CNRS, Universit\'e de Bourgogne, France}
\EmailD{\href{mailto:karol.kozlowski@u-bourgogne.fr}{karol.kozlowski@u-bourgogne.fr}}

\ArticleDates{Received November 28, 2013, in f\/inal form April 07, 2014; Published online April 11, 2014}

\Abstract{We establish several properties of the solutions to the linear integral equations describing the inf\/inite
volume properties of the XXZ spin-$1/2$ chain in the disordered regime.
In particular, we obtain lower and upper bounds for the dressed energy, dressed charge and density of Bethe roots.
Furthermore, we establish that given a~f\/ixed external magnetic f\/ield (or a~f\/ixed magnetization) there exists
a~unique value of the boundary of the Fermi zone.}


\Classification{45A13; 45M20}

\renewcommand{\thefootnote}{\arabic{footnote}} \setcounter{footnote}{0}

\section{Introduction}

The Bethe Ansatz constitutes a~powerful tool allowing one to map the problem of obtaining the spectrum of numerous
one-dimensional quantum Hamiltonians onto one of f\/inding solutions to a~system of algebraic equations, the so-called
Bethe equations.
The method was introduced in 1931 by H.~Bethe~\cite{BetheSolutionToXXX} with the example of the XXX spin-1/2 chain.
The latter corresponds to the $\Delta=1$ limit of the so-called XXZ spin-$1/2$ chain whose Hamiltonian reads
\begin{gather*}
H=J \sum\limits_{n=1}^{L} \big\{\sigma_n^x \sigma_{n+1}^{x}+\sigma_n^y \sigma_{n+1}^{y}+\Delta \sigma_n^z
\sigma_{n+1}^{z} \big\} -\frac{h}{2} \sum\limits_{n=1}^{L} \sigma_n^z.
\end{gather*}
Here $J>0$ is a~coupling constant measuring the strength of the exchange interaction, $\Delta$ is the longitudinal
anisotropy in the couplings, $h$ is an external magnetic f\/ield and the $\sigma_n^{a}$ are Pauli matrices acting
non-trivially on the $n^{\text{th}}$-quantum space $V_n \simeq \mathbb{C}^2$ in the tensor product decomposition
$\otimes_{n=1}^{L} V_n$ of the Hilbert space on which $H$ acts.

Although, in general, the Bethe equations, whose solutions provide a~set of quantum numbers parameterizing the
expectation values of observables of the f\/inite system, cannot be solved explicitly by analytic means (except at the
free fermion point $\Delta =0$), they are e.g.~the starting point for an exact analysis of the system in the
thermodynamic (inf\/inite volume) limit.
In this limit, based on still unproven but fairly reasonable assumptions, the Bethe Ansatz method enables an
ef\/f\/icient calculation of many of the physical observables of the model.
Quantities such as the total energy and momentum per-lattice site, the dressed energy and the momentum of the
excitations above the ground state, their dressed charge, etc.~-- the so-called thermodynamic functions~-- are directly
characterized by solutions to linear integral equations, which, for many models, take the form
\begin{gather}
f(\lambda)+\int_{-Q}^{Q} K(\lambda-\mu) f(\mu) \cdot \text{d} \mu=g(\lambda).
\label{ecriture forme general eqn int a etudier}
\end{gather}
The integral kernel $K$ and the domain of integration $[-Q;Q]$ are completely f\/ixed by the regime of the integrable
model, whereas the driving term $g$ depends on the specif\/ic thermodynamic quantity one is interested in.
Usually, one calls $g$ the bare thermodynamic function and $f$ its dressed counterpart.

The possibility to use linear integral equations to describe the thermodynamic limit was f\/irst observed by
Hulth\'en~\cite{HultenGSandEnergyForXXX} who built on it so as to propose an integral representation for the
ground-state energy per lattice site of the XXX spin-1/2 chain.
Then Orbach~\cite{OrbachXXZCBASolution} applied the Bethe Ansatz method to the spin-$1/2$ XXZ chain.
He generalized Hulth\'en's approach which allowed him to write down an integral representation for the ground-state
energy per lattice site of the XXZ spin-1/2 chain at $h=0$.
Orbach's integrand involved the solution to a~linear integral equation.
Since he was unable to solve it explicitly, he studied its solution numerically.
A year later, Walker~\cite{WalkerFirstExplicitSolToLinIntEqnMassiceXXZ} transformed Orbach's integral equation into one
of the type~\eqref{ecriture forme general eqn int a etudier} with $Q=\pi/2$ which then, using Fourier series, he was
able to solve explicitly.
Later, in 1964, Grif\/f\/iths~\cite{GriffithsXXZFirstLinIntEqnFiniteMagField} investigated the linear integral equations
associated with the XXZ spin-1/2 chain at non-vanishing magnetic f\/ield.
The above approaches focused on the evaluation of the ground state energy per-site.
In the mid-sixties, des Cloiseaux and Pearson~\cite{DescloizeauxPearsonExcitationsXXX} and des Cloiseaux and
Gaudin~\cite{DescloizeauxGaudinExcitationsXXZ+Gap} introduced linear integral equations that describe excitations above
the ground state, this at zero magnetic f\/ield.
All these works built on several assumptions on the large-volume behaviour of the Bethe roots describing the ground
state and low-lying excited states.
For the thermodynamic limit of the XXZ spin-1/2 chain at non-zero magnetic f\/ield Yang and
Yang~\cite{Yang-YangXXZStructureofGS} were able to prove some of these assumptions.
Still, it was only recently that Dorlas and
Samsonov~\cite{DorlasSamsonovThermoLim6VertexAndConvergceToDensityInSomeCases6VrtX} proved the condensation property of
the ground state Bethe roots for large-$L$ in the regime $-1<\Delta<0$ and for $h \not=0$.
Technicalities related with a~rigorous justif\/ication of its appearance set aside, the machinery of linear integral
equations appears utterly useful for understanding the thermodynamic limit of Bethe ansatz solvable models.
Over the years, it has been applied in numerous cases (see~\cite{BogoliubiovIzerginKorepinBookCorrFctAndABA} and
references therein).

The usefulness of linear integral equations goes, in fact, beyond the sole setting of a~model's thermodynamic limit.
Indeed, it is well known since the works~\cite{DeVegaWoynarowichFiniteSizeCorrections6VertexNLIEmethod,
GaudinTBAXXZMassiveInfiniteSetNLIE, KlumperBatchelorNLIEApproachFiniteSizeCorSpin1XXZIntroMethod,
TakahashiTBAforXXZFiniteTinfiniteNbrNLIE,Yang-YangNLSEThermodynamics} that one needs to recourse to non-linear integral
equations in order to describe integrable models of f\/inite size~$L$ or at f\/inite temperature $T$.
These are also useful to obtain various scaling limits towards massive and conformal quantum f\/ield
theories~\cite{BLZ97, BJMS10,DesVeg95,Zamolodchikov90}.
The coef\/f\/icients arising in the large-$L$ or low-$T$ asymptotic expansion of solutions to these non-linear integral
equations satisfy linear integral equations of the type~\eqref{ecriture forme general eqn int a etudier}.
In fact, our interest in the questions considered in this paper arose when we analysed the low-temperature limit of the
non-linear integral equations describing the spectrum of the quantum transfer matrix of the XXZ
chain~\cite{KozDugaveGohmannThermaxFormFactorsXXZ,DGK14app}.
We needed Proposition~\ref{Proposition Dressed energy positivity on other bank}
below in order to prove that the dressed energy provides a~lowest order low-temperature
approximation to the solution of the corresponding non-linear integral equation.

Although the existence and uniqueness of solutions to equations of the form $(I+K)\cdot f = g$ is usually easily
established using the general theory of linear integral equations, their expressions remain implicit.
The aim of the present paper is to prove certain overall properties of these solutions in the case of the linear
integral equations associated with the thermodynamic limit of the XXZ spin-$1/2$ chain in the disordered regime,
i.e.\ for $-1< \Delta < 1$.
In particular, we shall obtain explicit lower and upper bounds to the solutions $f$ for various driving terms~$g$ of interest.
We shall also discuss the construction of the endpoint of integration $Q$, and we shall show that the latter is well def\/ined.
Finally, to make this paper more self-contained, we shall as well review the calculation of the behaviour of solutions at large~$Q$.

It is worth mentioning that, for more complex quantum integrable models, the structure of the linear integral equation
changes~\cite{EsslerFrahmGohmanKlumperKorepinOneDimensionalHubbardModel, KulishReshetikhinNestedBAFirstIntroduction,
LiebWuFirstDerivationOfLIE4Hubbard1D} (the kernel $K$ may also depend on the sum of arguments or can be matrix valued).
We do stress, however, that, although we focus on a~specif\/ic model, the techniques we develop are general and also
applicable to other linear integral equations arising in the context of studying the ground-state properties of other
quantum integrable models.

The paper is organized as follows.
In Section~\ref{Section thermodynamic functions} we discuss the thermodynamic functions that we consider
(the dressed charge, the density of Bethe roots and the dressed energy) and give a~description of our results.
Then, in Section~\ref{Section Resolvent kernel}, we establish some properties of the resolvent kernel.
Finally, Sections~\ref{Section Dressed charge proofs},~\ref{Section dressed energy proofs}
and~\ref{Section dressed momentum proofs}
are devoted to the proofs of the results presented in Section~\ref{Section thermodynamic functions}.

\section{Functions characterizing the ground state of the XXZ chain}
\label{Section thermodynamic functions}

In the case of the XXZ spin-$1/2$ chain in the disordered regime, as brief\/ly outlined in the introduction, the
thermodynamic functions solve the linear integral equation
\begin{gather}
f(\lambda)+\int_{-Q_F}^{Q_F} K(\lambda-\mu|\gamma) f(\mu) \cdot \text{d} \mu= g(\lambda),
\label{ecriture eqn int generale}
\end{gather}
where $Q_F > 0$ is a~parameter called the Fermi rapidity and
\begin{gather*}
K (\lambda|\gamma)=\frac{\coth (\lambda-{\rm i}\gamma)-\coth (\lambda+{\rm i}\gamma)}{2 {\rm i}\pi} =
\frac{\sin(2\gamma)}{2\pi \sinh(\lambda+{\rm i}\gamma) \sinh(\lambda-{\rm i}\gamma)}
\end{gather*}
is the integral kernel depending parametrically on the variable $\gamma \in ]0;\pi[$ which parameterizes the
anisotropy parameter $\Delta = \cos(\gamma) \in {]{-}1;1[}$.
In the following, we shall often omit the auxiliary argument of the integral kernel and denote it simply by~$K(\lambda)$.
Therefore, throughout the paper, $K(\lambda)$ should \textit{always} be understood as $K(\lambda|\gamma)$.
The Fermi rapidity~$Q_F$ arising in~\eqref{ecriture eqn int generale} is f\/ixed by the external parameters of the
model, in particular by the magnetic f\/ield~$h$.
Its precise def\/inition will be given below.

Note also that there is a~qualitative change in the structure of the integral equation depending on the sign of
$\Delta$, since the integral kernel $K$ changes sign:
\begin{gather*}
K(\lambda|\gamma)>0
\quad
\text{for}
\quad
0< \gamma < \frac{\pi}{2}
\qquad
\text{and}
\qquad
K(\lambda|\gamma) < 0
\quad
\text{for}
\quad
\frac{\pi}{2} < \gamma < \pi.
\end{gather*}
Although, in the end, this has no big ef\/fect on the qualitative behaviour of the solutions, this fact has important
consequences for the techniques one needs to develop to establish the theorems presented below.
Finally, we recall that the kernel $K$ becomes trivial when $\gamma = \pi/2$.
This is the so-called free fermion point of the model, where dressed and bare quantities coincide.

We stress that the linear integral equations discussed above are all of truncated Wiener--Hopf type.
They can thus be solved perturbatively for large $Q$ by the Wiener--Hopf method.
The latter amounts to solving, asymptotically in $Q$, a~$2\times 2$ Riemann--Hilbert problem.
Furthermore, in the $Q=+\infty$ limit, the linear integral equations turn into convolution-type equations and, as such,
can be solved explicitly by means of a~Fourier transformation.

\subsection{The thermodynamic functions and the main results}

\subsubsection{The dressed charge}

The dressed charge $Z(\lambda|Q)$ is def\/ined as the solution to the integral equation
\begin{gather*}
Z(\lambda|Q)+\int_{-Q}^{Q} K(\lambda-\mu) \cdot Z(\mu|Q) \cdot \text{d} \mu=1.
\end{gather*}
It can be interpreted as the intrinsic moment of the magnetic excitations in the XXZ spin-$1/2$ chain.
Indeed, the thermodynamic limit of the average magnetization $\big\langle\sigma^z\big\rangle/2$ of a~state described by Bethe roots
that condense on $[-Q;Q]$ is given by
\begin{gather*}
\big\langle \sigma^z\big\rangle=1-2\int_{-Q}^{Q} Z(\lambda|Q) \cdot K(\lambda|\gamma/2)\cdot\text{d}\lambda.
\end{gather*}
The dressed charge is the thermodynamic function associated with the XXZ spin-$1/2$ chain which has the simplest
non-trivial driving term in~\eqref{ecriture eqn int generale}.
It is involved in the characterization of numerous observables in the XXZ chain.
For instance, when looking at the form of the low-lying excitations (of the order $1/L$) above the ground state in
f\/inite but large volume $L$, one obtains an expression of the
form~\cite{KlumperWehnerZittartzConformalSpectrumofXXZCritExp6Vertex, BogoliubiovIzerginKorepinBookCorrFctAndABA}
\begin{gather}
E_{\text{ex}}-E_{\text{G.S.}}
\simeq\frac{2\pi}{L} \cdot v_F \cdot\left\{\big(\ell \mathcal{Z}\big)^2+\left(\frac{s}{2\mathcal{Z}}\right)^2+n\right\},
\label{ecriture forme excitation basse energie}
\end{gather}
where $\ell$ is the Umklapp sector of the excitation, $s$ its spin and $n$ an integer.
The formula contains the constant $\mathcal{Z}=Z(Q_F|Q_F)$ which corresponds to the value taken by the dressed
charge at the Fermi rapidity $Q_F$ (see Section~\ref{Subsubsection dressed energy results}).
The above formula contains $v_F$ which has the interpretation of the velocity of excitations at $Q_F$.
We shall comment on this quantity later on.
The dressed charge also appears in the large-distance asymptotic behaviour of the longitudinal spin-spin correlation
functions~\cite{KozKitMailSlaTerXXZsgZsgZAsymptotics}
\begin{gather}
\big\langle \sigma_1^{z} \sigma_{m}^{z}\big\rangle=\big\langle \sigma^{z}\big\rangle^2-\frac{2\mathcal{Z}^2}{\pi^2 m^2}
+\sum\limits_{\ell \geq 1}{}
\big| \mathcal{F}\big[ \ell Z(*|Q)\big]\big|^2\cdot \frac{\cos(2\pi p_F m)}{m^{2\ell^2\mathcal{Z}^2}} \cdot
\left(1+O\left(\frac{\ln m}{m}\right)\right).
\label{ecriture formule asymptotiques grande distance 2 pts}
\end{gather}

In this formula $\mathcal{F}$ is a~certain explicit functional which represents the properly normalized form factor of
the spin operator taken between the model's ground state and an excited state with an $\ell$-Umklapp
excitation~\cite{KozKitMailSlaTerEffectiveFormFactorsForXXZ}.
The $*$ in~\eqref{ecriture formule asymptotiques grande distance 2 pts} indicates the running variable on which
$\mathcal{F}$ acts.
In the above asymptotic expansion, the constant $\mathcal{Z}$ parameterizes the magnitude of the oscillatory part of the
large-distance asymptotics.
In particular, depending on whether $\mathcal{Z}>1$ or $\mathcal{Z}<1$, the non-oscillatory asymptotics (which rather
ref\/lect ferromagnetic order) or the oscillatory ones (which ref\/lect the antiferromagnetic nature of the
interactions) will be dominant.
The constant~$p_F$ will be def\/ined below.

In order to state our results regarding the characterization of the dressed charge we need to introduce the domain
\begin{gather*}
\Upsilon_{\gamma}(Q)=\mathbb{C}\setminus\big\{\{[-Q;Q]+{\rm i}\gamma+{\rm i}n \pi:
n \in \mathbb{Z}\} \cup\{[-Q;Q]-{\rm i}\gamma+{\rm i}n \pi: n \in \mathbb{Z}\}\big\}.
\end{gather*}

\begin{theorem}
\label{Theorem pte dressed chge}
The dressed charge is a~smooth function of $(\lambda,Q) \in \mathbb{R} \times \mathbb{R}^+$ such that, pointwise in $Q$,
it is an ${\rm i}\pi$-periodic, holomorphic function of $ \lambda$ on $\Upsilon_{\gamma}(Q)$.
Furthermore, $\lambda \mapsto Z(\lambda|Q) $ is even and
\begin{itemize}
\item
for $0<\gamma < \pi/2$, is monotonically increasing on $\mathbb{R}^+$ and satisfies the bounds
\begin{gather}
\frac{1}{2\big(1- \gamma / \pi\big)} < Z(\lambda|Q) < 1;
\label{ecriture bornes pour Z ga low}
\end{gather}
\item
for $ \pi/2 < \gamma < \pi$, is monotonically decreasing on $\mathbb{R}^+$ and satisfies the bounds
\begin{gather}
1 < Z(\lambda|Q) < \frac{1}{2\big(1- \gamma / \pi\big)}.
\label{ecriture bornes pour Z ga high}
\end{gather}
\end{itemize}
Finally, one has the large-$Q$ behaviour
\begin{gather}
Z\big(Q|Q\big)=\sqrt{\frac{\pi}{2(\pi- \gamma)}}+O\big(e^{-2 Q \epsilon_{\gamma}}\big)
\qquad
\text{with}
\quad
\epsilon_{\gamma}=\min \left\{\frac{2\pi}{\pi-\gamma}, \frac{\pi}{\gamma}\right\}.
\label{ecriture DQ grand Q pour Z}
\end{gather}
\end{theorem}

\subsubsection{The density of Bethe roots and the dressed momentum}

Consider the sector of the XXZ spin-$1/2$ chain corresponding to the magnetization $ 1-2\mathfrak{m}_N$ with
$\mathfrak{m}_N = N/L$.
The ground state in this sector will be parameterized by $N$ Bethe roots $\lambda_1< \dots < \lambda_N$.
In the thermodynamic limit, $N, L \rightarrow+\infty$ with $\mathfrak{m}_N \rightarrow \mathfrak{m} \in [ 0; 1/2]$,
these Bethe roots will condense in an interval $[-Q_{\mathfrak{m}}; Q_{\mathfrak{m}} ]$ with a~density
$\rho(\lambda|Q_{\mathfrak{m}})$.
The function $\rho(\lambda|Q)$ solves the linear integral equation
\begin{gather}
\label{liebs}
\rho(\lambda|Q)+\int_{-Q}^{Q} K(\lambda-\mu) \cdot \rho(\mu|Q) \cdot \text{d}\mu= K(\lambda|\gamma/2).
\end{gather}
The endpoint $Q_{\mathfrak{m}}$ of the condensation interval, the so-called magnetic Fermi rapidity, is f\/ixed by the
equation
\begin{gather}
\int_{-Q_{\mathfrak{m}}}^{Q_{\mathfrak{m}}} \rho(\lambda|Q_{\mathfrak{m}}) \cdot \text{d}\lambda=\mathfrak{m}.
\label{ecriture equation definition endpoint Fermi zone dressed momentum}
\end{gather}

In order to af\/f\/irm that $\rho(\lambda|Q_{\mathfrak{m}})$ does indeed represent a~density, one should
establish, in particular, that it is a~positive function.
This property is relatively clear in the regime $\pi/2<\gamma<\pi$.
However, some tricks are necessary to establish it in the other regime $0<\gamma<\pi/2$.
Also, for various practical applications, it is useful to have explicit bounds on $\rho(\lambda|Q)$.
Finally, the density of Bethe roots is a~highly non-linear function of $Q$.
Therefore~\eqref{ecriture equation definition endpoint Fermi zone dressed momentum} is a~non-linear equation for the
magnetic Fermi rapidity.
It is not clear \textit{a priori} whether a~solution exists and, if yes, whether it is unique.
These properties follow from the theorem below.

\begin{theorem}
\label{Theorem pte densite}

The density of Bethe roots $\rho(\lambda|Q)$ is an ${\rm i}\pi$-periodic meromorphic function on~$\Upsilon_{\gamma}(Q)$ whose sole singularities are simple poles located at the points $ \pm {\rm i} \gamma / 2+ {\rm
i}n \pi $.
It is an even function of $\lambda$ and, uniformly in Q, is subject to the bounds
\begin{gather}
\rho_{\infty}(\lambda)<\rho(\lambda|Q)<K\big(\lambda|\gamma / 2\big)
\qquad
\text{for}
\ \
0<\gamma<\pi/2,
\label{ecriture bornes pour rho ga low}
\\
K\big(\lambda|\gamma / 2\big)<\rho(\lambda|Q)< \rho_{\infty}(\lambda)
\qquad
\text{for}
\ \
\pi/2<\gamma<\pi,
\label{ecriture bornes pour rho ga high}
\end{gather}
where
\begin{gather*}
\rho_{\infty}(\lambda)=\lim_{Q \rightarrow +\infty} \rho(\lambda|Q)=\left\{2\gamma \cosh\left(\frac{\pi\lambda}{\gamma}\right)\right\}^{-1}.
\end{gather*}
Furthermore, for any $\mathfrak{m} \in[0;1/2]$, there exists a~unique magnetic Fermi rapidity $Q_{\mathfrak{m}} \in
[0;+\infty[$ solving~\eqref{ecriture equation definition endpoint Fermi zone dressed momentum}.
The map $\mathfrak{m} \mapsto Q_{\mathfrak{m}}$ is smooth and monotonically increasing with derivative given by
\begin{gather*}
\partial_{\mathfrak{m}} Q_{\mathfrak{m}}
=\big\{2\rho\big(Q_{\mathfrak{m}}|Q_{\mathfrak{m}}\big) \cdot Z\big(Q_{\mathfrak{m}}|Q_{\mathfrak{m}}\big)\big\}^{-1}.
\end{gather*}
Finally, for $\pi/5<\gamma < \pi$ one has the large-$Q$ behaviour
\begin{gather*}
\rho(Q|Q)= e^{-\frac{\pi}{\gamma} Q} \cdot \sqrt{\frac{2}{\gamma}} \cdot\big(1-\gamma/\pi\big)^{\frac{\pi}{2\gamma}} \cdot
\frac{\Gamma\big(1+\pi/2\gamma\big)}{\Gamma((1+\pi/\gamma)/2)}
+O\big(e^{-2 Q \epsilon_{\gamma}}\big),
\end{gather*}
where $\epsilon_{\gamma}$ is as defined in~\eqref{ecriture DQ grand Q pour Z}.
\end{theorem}

The density of Bethe roots is connected with the so-called dressed momentum $p(\lambda|Q)$ in that
$\rho(\lambda|Q)$ corresponds to the density of momentum variables
\begin{gather}
p(\lambda|Q)=2\pi \cdot \int_{0}^{\lambda} \rho(\mu|Q) \cdot \text{d} \mu.
\label{definition integrale de p}
\end{gather}
One can also def\/ine the dressed momentum as the solution to the integral equation
\begin{gather}
p(\lambda|Q)+\int_{-Q}^{Q} K(\lambda-\mu) \cdot p(\mu|Q) \cdot \text{d} \mu=p_0(\lambda)+p\big(Q|Q\big) \cdot[ \theta(\lambda-Q)+\theta(\lambda+Q)],
\label{definition p via eqn lin int}
\end{gather}
where $p_0$ is the bare momentum and $\theta$ the bare phase
\begin{gather*}
p_0(\lambda)={\rm i}\ln\left(\frac{\sinh({\rm i}\gamma/2+\lambda)}{\sinh({\rm i}\gamma/2-\lambda)}\right)
\qquad
\text{and}
\qquad
\theta(\lambda)={\rm i}\ln\left(\frac{\sinh({\rm i}\gamma+\lambda)}{\sinh({\rm i}\gamma-\lambda)}\right).
\end{gather*}
Since $p_0^{\prime}(\lambda) = 2\pi K(\lambda|\gamma/2)$, it is readily seen that the
def\/inition~\eqref{definition p via eqn lin int} coincides with $\partial_{\lambda}p(\lambda|Q) = 2\pi
\rho (\lambda|Q ) $.
It should be also clear that, on the basis of~\eqref{definition integrale de p}, one can deduce the properties of the
dressed momentum from those of the density of Bethe roots.

\subsubsection{The dressed energy and the Fermi rapidity}
\label{Subsubsection dressed energy results}

The dressed energy $\varepsilon(\lambda|Q)$ satisf\/ies
\begin{gather}
\varepsilon(\lambda|Q) + \int_{-Q}^{Q} K(\lambda-\mu) \cdot \varepsilon(\mu|Q)
\cdot \text{d} \mu=\varepsilon_0(\lambda),
\label{Definition equation dressed charge}
\end{gather}
where $\varepsilon_0(\lambda)=h- 4\pi J \sin(\gamma) K(\lambda|\gamma/2)$ is the so-called bare energy.
In particular, $\varepsilon(\lambda|Q)$ is an implicit function of the external magnetic f\/ield $h>0$ and
of the magnitude of the exchange interaction $J$.

From the point of view of experiments on materials whose magnetic properties are modeled by the XXZ spin-1/2 chain, one
imposes an external magnetic f\/ield $h$.
The latter forces the magnetization of the material to adjust accordingly.
The XXZ-Hamiltonian can be diagonalized in every sector with a~f\/ixed magnetization.
Each of these sectors admits a~f\/ixed magnetization ground state which corresponds to a~condensation of Bethe roots in
the interval $[-Q_{\mathfrak{m}}; Q_{\mathfrak{m}} ]$.
Then the overall ground state of the model is to be chosen among all these f\/ixed magnetization ground states in such
a~way that the interaction with the external magnetic f\/ield leads to the lowest possible energy.
This means that one should choose the Fermi sea of the model $[-Q_F;Q_F]$ in such a~way that the energy of excitations
exactly at the Fermi boundary $Q_F$ vanishes and is negative inside of $[-Q_F;Q_F]$ and positive outside,  viz.\
on $\mathbb{R} \setminus [-Q_F;Q_F]$.
The reason for this is that the ground state built in such a~way has the property that creating a~hole inside the Fermi
zone or adding a~particle outside of it necessarily increases the energy.
In other words, the Fermi rapidity is def\/ined by the equation
\begin{gather*}
\varepsilon(Q_F|Q_F)=0.
\end{gather*}
Note that, just as $\rho(\lambda|Q)$, $\varepsilon(\lambda|Q)$ depends parametrically on $Q$ in a~complicated
way.
Thus, again, it is not clear \textit{a priori} whether the above equation does admit solutions at all and, if, yes
whether these are unique or not.

Within this context, the so-called particle-hole excitations above the ground state take the form
\begin{gather*}
E_{\text{ex}}-E_{\text{G.S.}}= \sum\limits_{a=1}^{n_p} \varepsilon\big(\lambda_a^{(p)}\big|Q_F\big)-
\sum\limits_{a=1}^{n_h} \varepsilon\big(\lambda_a^{(h)}\big|Q_F\big)
\end{gather*}
in which $\lambda_a^{(p)} \in \mathbb{R} \setminus [-Q_F;Q_F]$ are the particle rapidities and $\lambda_a^{(h)} \in
[-Q_F;Q_F]$ are the hole rapidities.

At this stage of our discussion we are f\/inally in position to def\/ine of the Fermi velocity which appeared
in~\eqref{ecriture forme excitation basse energie}.
In terms of the dressed energy and the momentum of excitations it is given as
\begin{gather*}
v_F=\frac{\partial_{\lambda} \varepsilon(Q_F|Q_F)}{\partial_{\lambda} p(Q_F|Q_F)}.
\end{gather*}
Here,
$\partial_{\lambda}$ refers to the derivative in respect to the f\/irst argument of the functions.
To close, the constant $p_F$ governing the speed of the oscillations in~\eqref{ecriture formule asymptotiques grande
distance 2 pts} corresponds to the value of the dressed momentum on the Fermi boundary $Q_F$,  viz.\
$p(Q_F|Q_F) = p_F$.

\begin{theorem}
\label{Theorem pte energie habillee}
The dressed energy is a~smooth function of $(\lambda,Q)\in \mathbb{R} \times [0;+\infty[$.
Furthermore, pointwise in $Q$, it is an ${\rm i}\pi$-periodic meromorphic function of $\lambda$ on
$\Upsilon_{\gamma}(Q)$ with simple poles at $\lambda = \pm {\rm i}\gamma / 2+{\rm i}n \pi $.

$\varepsilon(\lambda|Q)$ satisfies the bounds
\begin{gather*}
\forall\, Q \leq Q_0
\qquad
 \begin{cases}
\widetilde{\varepsilon}\big(\lambda\big)>\varepsilon(\lambda|Q)> \varepsilon_0(\lambda) &\text{for} \ \ 0 < \gamma < \pi/2,
\\
\varepsilon_0(\lambda)>\varepsilon(\lambda|Q)>\widetilde{\varepsilon}\big(\lambda\big) &\text{for} \ \ \pi/2 < \gamma < \pi,
\end{cases}
\end{gather*}
where $Q_0$ corresponds to the unique positive zero of $\varepsilon_0(\lambda)$ and the function
$\widetilde{\varepsilon}$ is given by
\begin{gather*}
\widetilde{\varepsilon}(\lambda)= h-\frac{2\pi J \sin (\gamma)} {\gamma \cosh\big(\frac{\pi}{\gamma} \lambda\big)}.
\end{gather*}
Further, the bound involving $\widetilde{\varepsilon}$ holds, in fact, for any value of $Q$.

For any $h>0$, there exists a~unique solution $Q_F>0$ to the equation $\varepsilon(Q_F|Q_F)=0$.
Let $\widetilde{Q}>0$ be the unique positive zero of $\widetilde{\varepsilon}(\lambda)$.
Then the Fermi rapidity $Q_F$ satisfies the bounds
\begin{gather*}
\widetilde{Q} < Q_F < Q_0
\quad
\text{for}
\quad
0<\gamma<\pi/2\qquad
\text{and}
\qquad
\widetilde{Q}>Q_F>Q_0
\quad
\text{for}
\quad
\pi/2<\gamma<\pi.
\end{gather*}
The function $h \mapsto Q_F$ is a~smooth, monotonically decreasing, function of $h$ with $h$-derivative given by
\begin{gather*}
\partial_{h} Q_F=- \frac{Z(Q_F|Q_F)}{\partial_{\lambda} \varepsilon(Q_F|Q_F)}.
\end{gather*}
For $\pi > \gamma > \pi/5$ it has the small-$h$ asymptotic behaviour given by
\begin{gather*}
\exp\left(\frac{\pi}{\gamma} Q_F(h)\right)=\frac{8\pi J \sin(\gamma)}{\sqrt{\gamma} \cdot h} \cdot
\left(1-\frac{\gamma}{\pi}\right)^{\frac{\pi+\gamma}{2\gamma}} \cdot \frac{\Gamma\big(1+\pi/2\gamma\big)}{\Gamma((1+\pi/\gamma)/2)}\cdot(1+o(1)).
\end{gather*}
\end{theorem}

In fact, the dressed energy evaluated precisely at the Fermi rapidity $Q_F$ has several very natural properties.
In particular, its real part is positive on the line $\mathbb{R}-{\rm i}\gamma$.
This property is crucial for a~consistency check of the low-$T$ asymptotic
expansion~\cite{KozDugaveGohmannThermaxFormFactorsXXZ} of the solution to the non-linear integral
equation~\cite{KlumperNLIEfromQTMDescrThermoXYZOneUnknownFcton} driving the f\/inite temperature properties of the XXZ
spin-$1/2$ chain.

\begin{proposition}\label{Proposition Dressed energy positivity on other bank}
The dressed energy evaluated at the Fermi rapidity $\lambda \mapsto \varepsilon(\lambda|Q_F)$ is a~monotonically
increasing function on $\mathbb{R}^+$ with a~unique zero at $Q_F$ that satisfies
\begin{gather}
\Re\big[ \varepsilon_+(\lambda-{\rm i}\gamma|Q_F)\big]>\frac{h}{4}
\qquad
\text{for all}
\quad
\lambda \in \mathbb{R}
\qquad
\text{and}
\qquad
0<\gamma<\pi/2.
\label{Ecriture lower bound on other bank}
\end{gather}
Here, $\varepsilon_+$ refers to the boundary value of $\varepsilon$ when approaching a~point on $\mathbb{R}-{\rm i}\gamma$ from above.
\end{proposition}

\section{The resolvent kernel}
\label{Section Resolvent kernel}

The resolvent kernel $R_{Q}(\lambda,\mu)$ is def\/ined as the integral kernel of the inverse operator $I-R_Q$ to $I+K$
understood as acting on $\mathcal{C}^{0}([-Q;Q])$, the space of continuous functions on the interval $
[-Q;Q]$.
It satisf\/ies the integral equation
\begin{gather*}
R_{Q}(\lambda,\mu)+\int_{-Q}^{Q} K(\lambda-\nu) \cdot R_{Q}(\nu,\mu) \cdot \text{d} \nu= K(\lambda-\mu).
\end{gather*}
Using the resolvent kernel the solution to~\eqref{ecriture eqn int generale} can be represented as
\begin{gather*}
f(\lambda)=g(\lambda)-\int_{-Q}^{Q} R_{Q}(\lambda,\mu) \cdot g(\mu) \cdot \text{d} \mu.
\end{gather*}
$R_Q$ is thus the most important object associated with this integral equation.
In the present section, we establish its existence (i.e.~the invertibility of $I+K$) and prove several of its properties.

Also, since the kernel $K$ depends on the dif\/ference of its arguments, the integral equations are of truncated
Wiener--Hopf type.
As such, they are exactly solvable in the $Q \rightarrow +\infty$ limit by means of Fourier transformation.
In this limit, the Neumann series def\/ines a~kernel $R$ solely depending on the dif\/ference of variables $\lambda-\mu$
which solves the convolution-type integral equation
\begin{gather*}
R(\lambda-\mu)+\int_{\mathbb{R}}{} K(\lambda-\nu) \cdot R(\nu-\mu) \cdot \text{d} \nu= K(\lambda-\mu).
\end{gather*}

\subsection{Overall properties and bounds}

\begin{proposition}
The operator $I+K:
\mathcal{C}^{0}([-Q;Q]) \mapsto \mathcal{C}^{0}\big([-Q;Q]\big)$ is invertible,
and its resolvent kernel is well defined and given in terms of the Neumann series
\begin{gather*}
R_{Q}(\lambda,\mu)=K(\lambda-\mu) -\sum\limits_{n \geq 1}{}\! (-1)^{n-1} \!\int_{-Q}^{Q}\! K(\lambda-\nu_1) \cdot
\prod\limits_{a=1}^{n-1}\big\{K(\nu_{a}-\nu_{a+1})\big\} \cdot K(\nu_n-\mu) \cdot {\rm d}^n \nu.
\end{gather*}
The resolvent kernel $R_{Q}(\lambda,\mu)$ is a~symmetric function of $(\lambda,\mu)$.
It is also a~smooth function in $(\lambda,\mu,Q) \in \mathbb{R}^2\times \mathbb{R}^+$.
\end{proposition}

\begin{proof}
The Neumann series is convergent in the sup-norm topology $||f||_{\infty} = \max\big\{|f(\mu)|:\mu \in[-Q;Q]\big\}$ since,
for any $f\in \mathcal{C}^{0}([-Q;Q])$ such that $||f||_{\infty} \leq 1$ one has
\begin{gather}
|| K. f ||_{\infty} \leq ||f||_{\infty} \cdot \max_{\lambda \in [-Q;Q]} \int_{\mathbb{R}}{}\big|
K(\lambda- \mu)\big| \cdot \text{d} \mu \leq\left| 1- 2\frac{\gamma}{\pi}\right| < 1.
\label{Bound on norm K}
\end{gather}
The latter implies the uniform convergence of the Neumann series with exponential speed.
The bound~\eqref{Bound on norm K} also ensures that the spectral radius of the integral operator $K$ is bounded by
$\big| 1- 2\frac{\gamma}{\pi}\big|$.
Moreover, using standard dif\/ferentiation under the integral sign theorems, it follows that $R_{Q}(\lambda,\mu)$ is
smooth in $(\lambda,\mu,Q) \in \mathbb{R}^2\times \mathbb{R}^+$.
\end{proof}

We now focus on the $Q=+\infty$ case.
We establish some of the most fundamental properties of the resolvent $R$ at $Q=+\infty$ that will be most useful for
the analysis that shall follow:
\begin{lemma}\label{Lemme: propr R}\quad
\begin{enumerate}[$(i)$]\itemsep=0pt
\item
$R$ has the Fourier integral representation
\begin{gather}
\label{fintr}
R (\lambda)=\int_{\mathbb{R}}{} \frac{\sinh\big[\big(\pi/ 2-\gamma\big)k\big] e^{- {\rm i}k \lambda}}
{\cosh (\gamma k/ 2) \sinh[(\pi/2-\gamma/2)k]} \frac{{\rm d} k}{4 \pi}.
\end{gather}
\item
For $0<\gamma<\pi/2$, $R$ is even and positive on $\mathbb{R}$, monotonically decreasing on $\mathbb{R}^+$ and
satisfies $\lim\limits_{\lambda \rightarrow \infty} R (\lambda) = 0$.
\end{enumerate}
\end{lemma}

\begin{proof}
Item (i) follows from the convolution theorem using
\begin{gather}
\mathcal{F}[K](k)=\frac{\sinh\big[\big(\frac{\pi}{2}-\gamma\big)k\big]} {\sinh\big(\frac {\pi k}2\big)},
\qquad
1+\mathcal{F}[K](k)=\frac{2\cosh\big(\frac{\gamma k} 2\big) \sinh\big[\big(\frac\pi 2-\frac \gamma 2\big)k\big]}
{\sinh\big(\frac {\pi k}2\big)},
\label{Ecriture Fourier K et 1+K}
\end{gather}
where
\begin{gather*}
\mathcal{F}[g](k)=\int_{\mathbb{R}}{} g(\lambda) e^{{\rm i}k\lambda} \cdot \text{d} \lambda.
\end{gather*}
In order to prove (ii) observe that
\begin{gather*}
\mathcal{F}[g](k)=\frac{\gamma \pi}{\pi-\gamma} \cdot \frac{1}{\cosh\big[ \frac{\gamma \pi k}{2(\pi -\gamma)}\big]},
\qquad \text{where} \qquad g(\lambda)=\frac{1}{\cosh\big[(1-\pi/\gamma) \lambda\big]}.
\end{gather*}
This allows one to recast $R$, for $0 < \gamma < \pi/2$, as
\begin{gather}
\label{convreff}
R (\lambda)=\frac{\pi}{2\gamma(\pi-\gamma)} \int_{\mathbb{R}}{} \frac{K\Big(\frac{\mu}{1-\gamma/ \pi}
|\gamma^{\prime}\Big)}{\cosh [\pi(\lambda-\mu)/\gamma]} \cdot \text{d} \mu > 0,
\qquad
\text{where}
\qquad
\gamma'=\frac{\gamma/2}{1-\gamma/ \pi}
\end{gather}
(compare~\cite[Appendix~C]{Yang-YangXXZStructureofGS}).
It is clear from the above representation that $R$ is even, asympto\-ti\-cally zero, and that it is monotonically decreasing
for $\lambda > 0$.
\end{proof}

The function $R$ is one of the most prominent functions in the theory of the XXZ model.
It appears as the logarithmic derivative of the two-spinon scattering phase and determines the free energy of the
six-vertex model in the critical regime.
It satisf\/ies nice functional equations and can be expressed in terms of Barnes double-gamma functions.

In order to be able to extract valuable informations out of the resolvent kernel $R_{Q}(\lambda,\mu)$ one needs to
obtain lower (or upper) bounds on it.
This is the purpose of the lemma below.
\begin{lemma}
\label{Lemme bounds up and low resolvent}
The resolvent kernel satisfies the bounds
\begin{gather}
R_{Q}(\lambda,\mu)>R(\lambda-\mu) \qquad\text{for}\quad  0<\gamma<\pi/2,
\nonumber
\\
R(\lambda-\mu) < R_{Q}(\lambda,\mu) < 0 \qquad\text{for}\quad  \pi/2<\gamma<\pi
\label{Ecriture bornes sup et inf pour resolvent en termes gamma}
\end{gather}
uniformly in $(\lambda,\mu) \in \mathbb{R}^2$.
Furthermore, for $\lambda, \mu >0$, it also satisfies the inequalities
\begin{gather}
R_{Q}(\lambda,\mu)-R_{Q}(\lambda,-\mu)>0 \qquad\text{for}\quad  0<\gamma<\frac{\pi}{2},
\nonumber
\\
R_{Q}(\lambda,\mu)-R_{Q}(\lambda,-\mu) < 0 \qquad\text{for}\quad  \frac{\pi}{2}<\gamma<\pi.
\label{Ecriture comparaison resolvent en mu et moins mu}
\end{gather}
\end{lemma}
A quite non-trivial consequence of the above lemma is that $R_{Q}(\lambda,\mu)$ is a~strictly positive function when $0<\gamma<\pi/2$.
This statement is highly non-trivial while solely looking at the Neumann series for this operator.

\begin{proof}
The bound for $\pi/2<\gamma<\pi$ is readily deduced by term-wise majorations in the Neumann series, since the series for
$-R_{Q}(\lambda,\mu)$ is a~sum of strictly positive terms.
In order to establish the second bound, we recast the integral equation satisf\/ied by $R_{Q}(\lambda, \mu)$ in the form
\begin{gather*}
R_Q(\lambda,\mu)+\int_{\mathbb{R}}{}K(\lambda-\nu) R_{Q}(\nu,\mu) \cdot \text{d} \nu
-\int_{\mathbb{R} \setminus [-Q;Q]}{} K(\lambda-\nu) R_{Q}(\nu,\mu) \cdot
\text{d} \nu=K(\lambda-\mu).
\end{gather*}
Then, acting with the inverse operator $I-R$ on $I+K$ understood as an integral operator on~$\mathcal{C}_b^{0}(\mathbb{R})$, the space of continuous functions on $\mathbb{R}$ that are bounded at inf\/inity,
recasts the above integral equation as
\begin{gather*}
R_Q(\lambda,\mu)-\int_{\mathbb{R} \setminus [-Q;Q]}{} R(\lambda-\nu) \cdot
R_{Q}(\nu,\mu) \cdot \text{d} \nu=R(\lambda-\mu).
\end{gather*}
The Neumann series for the resolvent $\mathcal{R}$ of the integral operator $I-R$ understood as acting on
$\mathcal{C}_b^{0}(\mathbb{R}\setminus[-Q;Q])$ converges uniformly, for the very same reasons that were
invoked for the operator $K$ acting on $ \mathcal{C}_b^{0}([-Q;Q])$, since, when $0<\gamma< \pi/2$,
cf.~\eqref{fintr},
\begin{gather*}
\int_{\mathbb R}  R(\lambda) \cdot \text{d} \lambda = 1-\frac{\pi}{2(\pi-\gamma)} < 1.
\end{gather*}
Since this Neumann series for $\mathcal{R}$ is a~sum of strictly positive terms it follows that $R_{Q}(\lambda,\mu) >
R(\lambda-\mu)$.
The bounds~\eqref{Ecriture comparaison resolvent en mu et moins mu} at $\pi/2<\gamma<\pi$ follow from the series of
multiple integral representations
\begin{gather*}
R_{Q}(\lambda,\mu)- R_Q(\lambda,-\mu) = K(\lambda-\mu) - K(\lambda+\mu)
+\sum\limits_{n \geq 1}{} \int_{0}^{Q}\big[ K(\lambda- \tau_{1})-K(\lambda+\tau_{1})\big]
\\
\qquad{}
\times
\prod\limits_{\ell =1}^{n-1}\big[ K(\tau_{\ell}- \tau_{\ell +1})-K(\tau_{\ell}+\tau_{\ell +1})\big] \cdot\big[
K(\tau_{n}-\mu)-K(\tau_{n}+\mu)\big] \cdot \text{d}^n \tau
\end{gather*}
and an analogous representation involving $R$ for $0<\gamma < \pi/2$.
\end{proof}

\subsection[Asymptotic representation at large $Q$]{Asymptotic representation at large $\boldsymbol{Q}$}

The resolvent kernel $R_{Q}(\lambda,\mu)$ of the operator $I+K$ at large $Q$ can be constructed by means of an
asymptotic resolution of a~$2\times 2$ Riemann--Hilbert problem.
This has been carried out in~\cite{KozKitMailSlaTerRHPapproachtoSuperSineKernel}.
We recall these results here.
The resolvent $R_{Q}(\lambda,\mu)$ can be decomposed as
\begin{gather}
R_{Q}(\lambda,\mu)=R_{Q}^{(0)}(\lambda,\mu)+R_{Q}^{(\text{pert})}(\lambda,\mu),
\label{decomposition RQ principal et pert}
\end{gather}
where
\begin{gather}
R_{Q}^{(0)}(\lambda,\mu)=\int_{\mathbb{R}}{} \frac{\text{d} \xi \text{d} \eta}{4{\rm i}\pi^2}
\mathcal{F}[K](\xi) \cdot\left\{\frac{\alpha_+(\eta)}{\alpha_-(\xi)} e^{{\rm i}Q(\xi-\eta)}
-\frac{\alpha_+(\xi)}{\alpha_-(\eta)} e^{-{\rm i}Q(\xi-\eta)}\right\} \cdot \frac{ e^{{\rm i}(\mu
\eta-\lambda \xi)}}{\xi-\eta}
\label{ecriture R ppl}
\end{gather}
and, uniformly in $(\lambda,\mu)$, one has
\begin{gather*}
\big| R_{Q}^{(\text{pert})}(\lambda,\mu)\big| \leq C e^{-2\epsilon_{\gamma} Q}
\qquad
\text{with}
\quad
\epsilon_{\gamma}=\text{min}\left\{\frac{2\pi}{\pi-\gamma},\frac{\pi}{\gamma}\right\}.
\end{gather*}
Note that $\epsilon_{\gamma}$ corresponds to the imaginary part of the zero of the function $1+ \mathcal{F}[K]$ closest
to the real axis.
The function $\alpha$ is analytic on $ \mathbb{C}\setminus \mathbb{R}$, goes to $1$ at $\infty$ and satisf\/ies the jump
condition on $\mathbb{R}$
\begin{gather*}
\alpha_{+}(\xi) \cdot\big(1+\mathcal{F}[K](\xi)\big)= \alpha_{-}(\xi).
\end{gather*}
Above, $ \alpha_{\pm}(s)$ correspond to the boundary values when $z$ approaches a~point $s \in \mathbb{R}$ from the
upper ($+$) or lower ($-$) half-plane.
In fact, $\alpha$ can be expressed in terms of $\Gamma$-functions as
\begin{gather*}
\alpha(\lambda)=\sqrt{2(\pi-\gamma)}\cdot\frac{(1-\gamma/\pi)^{\frac{{\rm i}\lambda}{2}(1-\frac{\gamma}{\pi})}\cdot
(\gamma/\pi)^{\frac{{\rm i}\lambda \gamma}{2\pi}}\cdot\Gamma(1+ {\rm i}\lambda/2)}{\Gamma\Big(\frac{1+{\rm i}\lambda\gamma/\pi}{2}\Big)\cdot
\Gamma(1+{\rm i}\frac{\lambda}{2} (1-\frac{\gamma}{\pi}))}
\qquad
\text{for}
\quad
\Im(\lambda)< 0,
\end{gather*}
and one has the symmetry $\alpha(\lambda)=1/\alpha(-\lambda)$, which yields the expression for $\alpha$ in the
upper-half plane.

\section{The dressed charge}\label{Section Dressed charge proofs}

\subsection{General bounds}

The dressed charge can be explicitly expressed in terms of the resolvent kernel as
\begin{gather*}
Z\big(\lambda|Q\big)=1- \int_{-Q}^{Q} R_{Q}(\lambda,\mu) \cdot \text{d} \mu.
\end{gather*}
Thus, it follows from the bounds~\eqref{Ecriture bornes sup et inf pour resolvent en termes gamma} for $\pi/2<\gamma <
\pi$, that
\begin{gather*}
1<Z\big(\lambda|Q\big)<1- \int_{-Q}^{Q} R(\lambda-\mu) \cdot \text{d} \mu< 1- \int_{\mathbb{R}}{}
R(\mu) \cdot \text{d} \mu=\frac{\pi}{2(\pi-\gamma)}.
\end{gather*}
Since $R(\lambda)>0$ when $ 0 <\gamma < \pi/2 $ due to~\eqref{Ecriture bornes sup et inf pour resolvent en termes
gamma}, one obtains that $R_{Q}(\lambda,\mu)>0$ which implies that $Z(\lambda|Q)<1$ in this regime.
Further, repeating the change of integration contour trick, we get that $Z(\lambda|Q)$ solves
\begin{gather*}
Z(\lambda|Q)=\frac{\pi}{2 (\pi- \gamma)}+\int_{\mathbb{R} \setminus [-Q;Q]}{} R(\lambda-\mu) Z(\mu|Q)\cdot\text{d}\mu.
\end{gather*}
Using that the resolvent $\mathcal{R}(\lambda,\mu)$ to $I-R$ is positive, we get the lower bound.

It follows from the smoothness properties of the resolvent kernel $R_{Q}(\lambda,\mu)$ and the compactness of $[-Q;Q]$
that $Z(\lambda|Q)$ is a~smooth function of $(\lambda, Q) \in \mathbb{R} \times \mathbb{R}^+$.
Then, dif\/ferentiation under the integral sign and integration by parts implies that $\partial_{\lambda}Z$ solves the
integral equation
\begin{gather*}
\partial_{\lambda}Z(\lambda|Q)+\int_{-Q}^{Q}K(\lambda-\mu) \cdot \partial_{\mu}Z(\mu|Q)\cdot\text{d}\mu
=[K(\lambda-Q)-K(\lambda+Q)] \cdot Z(Q|Q).
\end{gather*}
Thus,
\begin{gather*}
\partial_{\lambda}Z(\lambda|Q)=[ R_{Q}(\lambda,Q)-R_{Q}(\lambda,-Q)] \cdot Z(Q|Q),
\end{gather*}
so that one can conclude about the strict monotonicity of $\lambda \mapsto Z(\lambda|Q)$ in virtue of
equation~\eqref{Ecriture comparaison resolvent en mu et moins mu}.

\subsection[Behaviour of $Z(Q|Q)$ at large $Q$]{Behaviour of $\boldsymbol{Z(Q|Q)}$ at large $\boldsymbol{Q}$}

We now establish the large-$Q$ behaviour by following the steps in~\cite{KozKitMailSlaTerRHPapproachtoSuperSineKernel}.
Due to~\eqref{decomposition RQ principal et pert}, \eqref{ecriture R ppl}
\begin{gather*}
Z(Q|Q)-1=-\int_{-Q}^{Q} R_{Q}(Q,\mu) \cdot \text{d} \mu= \mathcal{V}_Q+ O\big(Q
 e^{-2\epsilon_{\gamma} Q}\big),
\end{gather*}
where we have used the explicit bounds on $R_{Q}^{(\text{pert})}$ and have set
\begin{gather*}
\mathcal{V}_Q=- \int_{\mathbb{R}}{} \frac{\text{d} \xi \text{d} \eta}{4{\rm i}\pi^2}\cdot \frac{
\mathcal{F}[K](\xi)}{\xi-\eta}\left\{\frac{\alpha_+(\eta)}{\alpha_-(\xi)} e^{- {\rm
i}Q\eta}-\frac{\alpha_+(\xi)}{\alpha_-(\eta)} e^{{\rm i}Q\eta -2 {\rm i}\xi Q}\right\} \cdot
\frac{ e^{{\rm i}\eta Q}- e^{-{\rm i}\eta Q}}{{\rm i} \eta}.
\end{gather*}
We now carry out the large-$Q$ asymptotic expansion of the above expression.
We deform the $\eta$-integration contour to $\mathbb{R}+i\kappa $ for some $\kappa >0$ small enough and use the jump
condition satisf\/ied by $\alpha$.
This yields
\begin{gather*}
\begin{split}
&\mathcal{V}_Q=- \int_{\mathbb{R}+{\rm i}\kappa}{} \frac{\text{d} \eta}{2 {\rm i}\pi}\int_{\mathbb{R}}{} \frac{\text{d} \xi}{2 {\rm i}\pi}
\left\{\alpha_{+}(\eta) \cdot \frac{\alpha_+^{-1}(\xi)-\alpha_-^{-1}(\xi)}{(\xi-\eta)\eta}\big(1- e^{-2 {\rm i}Q\eta}\big)\right.
\\
& \left.\phantom{\mathcal{V}_Q=}{}
- e^{-2{\rm i}Q\xi} \frac{\alpha_{-}(\xi)-\alpha_{+}(\xi)}{\alpha_{-}(\eta) (\xi-\eta)\eta}\big(e^{2 {\rm i}Q\eta}-1\big)\right\}.
\end{split}
\end{gather*}
We now decompose $\alpha_+^{-1}(\xi)-\alpha_-^{-1}(\xi) = \alpha_+^{-1}(\xi) -1+1-\alpha_-^{-1}(\xi) $
(resp.~$\alpha_{-}(\xi)-\alpha_{+}(\xi)= \alpha_{-}(\xi) -1+1-\alpha_{+}(\xi)$) in the f\/irst (resp.\
second) term.
By deforming the $\xi$-integration to $-i\infty$ one can then drop all the contribution of $1-\alpha_-^{-1}(\xi)$
(resp.
$\alpha_{-}(\xi) -1 $) leading to
\begin{gather*}
\mathcal{V}_Q=- \int_{\mathbb{R}+{\rm i}\kappa}{}
\frac{\text{d} \eta}{2 {\rm i}\pi} \int_{\mathbb{R}}{}
\frac{\text{d} \xi}{2 {\rm i}\pi}\left\{\alpha_{+}(\eta) \cdot \frac{\alpha_+^{-1}(\xi)-1}{(\xi-\eta)\eta}
\big(1- e^{-2 {\rm i}Q\eta}\big)\right\}
\\
\phantom{\mathcal{V}_Q=}{}
+ \int_{\mathbb{R}+{\rm i}\kappa}{} \frac{\text{d} \eta}{2 {\rm i}\pi} \int_{
\mathbb{R}-{\rm i}\kappa}{} \frac{\text{d} \xi}{2 {\rm i}\pi}\left\{ e^{-2 {\rm i}Q\xi}
\frac{1-\alpha_{+}(\xi)}{\alpha_{-}(\eta) (\xi-\eta)\eta}\big(e^{2 {\rm i}Q\eta}-1\big)\right\}.
\end{gather*}
The second line is already $O\big(e^{-2\kappa Q}\big)$.
The contribution of the f\/irst line can be obtained by deforming the $\xi$-integration to $+{\rm i}\infty$ and picking
up the pole at $\xi=\eta$.
Thus,
\begin{gather*}
\mathcal{V}_Q=- \int_{\mathbb{R}+{\rm i}\kappa}{}
\frac{\text{d} \eta}{2 {\rm i}\pi} \frac{1- \alpha_+(\eta)}{\eta}\big(1- e^{-2 {\rm i}Q\eta}\big) +O\big(e^{-2\kappa Q}\big)
\\
\phantom{\mathcal{V}_Q}{}
= \alpha_+(0) -1+ \int_{\mathbb{R}-{\rm i}\kappa}{} \frac{1- \alpha_+(\eta)}{\eta} e^{-2 {\rm
i}Q\eta} \cdot \frac{\text{d} \eta}{2 {\rm i}\pi} +O\big(e^{-2\kappa Q}\big).
\end{gather*}
The integral term, again, is $O\big(e^{-2\kappa Q}\big)$.
Then, it follows from $\alpha_+(\lambda) = \alpha_-^{-1}(-\lambda)$ that
\begin{gather*}
\alpha_+^2(0)=\frac{1}{1+ \mathcal{F}[K ](0)}=\frac{\pi}{2(\pi-\gamma)}.
\end{gather*}
In other words, one has
\begin{gather*}
Z(Q|Q)=\sqrt{\frac{\pi}{2(\pi-\gamma)}}+O\big(Q e^{-2\epsilon_{\gamma}Q}\big)
\end{gather*}
whence the optimal value for $\kappa$ is $\epsilon_{\gamma}$.

\section{The density of Bethe roots}\label{Section dressed momentum proofs}

\subsection{General bounds}

In this section we establish Theorem~\ref{Theorem pte densite}.
The meromorphicity and location of cuts is readily deduced from the linear integral equation def\/ining
$\rho(\lambda|Q)$.
In order to establish the lower and upper bounds it is enough to repeat the strategy that has already been employed for
the dressed charge and use that
\begin{gather*}
\frac{p_0^{\prime}(\lambda)}{2\pi}-\int_{\mathbb{R}}{} R(\lambda-\mu) \cdot p_0^{\prime} (\mu) \cdot \frac{\text{d} \mu}{2\pi} =\rho_{\infty}(\lambda).
\end{gather*}
Thus, it remains to establish the statement relative to the magnetic Fermi rapidity $ Q_{\mathfrak{m}}$.

It follows from a~dif\/ferentiation of the linear integral equation satisf\/ied by the density of Bethe roots that
\begin{gather*}
\partial_{Q}\rho(\lambda|Q)=-\rho(Q|Q)\cdot[R_{Q}(\lambda,Q)+R_{Q}(\lambda,-Q)].
\end{gather*}
Hence, taking into account the smoothness of $\rho(\lambda|Q)$ in $(\lambda,Q)$, the function
\begin{gather*}
f(Q)= \int_{-Q}^{Q} \rho\big(\lambda|Q\big) \cdot \text{d} \lambda
\end{gather*}
is smooth and
\begin{gather*}
f^{\prime}(Q)=2\rho(Q|Q) \cdot \left\{1-\int_{-Q}^{Q} R_{Q}(\lambda,Q) \cdot \text{d}\lambda \right\} =2\rho(Q|Q) \cdot \mathcal{Z}(Q|Q)> 0.
\end{gather*}
$f(Q)$ is thus a~strictly increasing function of $Q$.
Clearly $f(0)=0$.
Moreover, using the explicit solution of Lieb's equation~\eqref{liebs} for $Q=+\infty$, we f\/ind that $f(+\infty)=1/2$.
$f$ is thus a~dif\/feomorphism from $[0;+\infty[$ onto $[0;1/2]$.
This proves the uniqueness and existence of solutions to~\eqref{ecriture equation definition endpoint Fermi zone dressed momentum}
along with the smoothness in $\mathfrak{m}$ of the solution $Q_{\mathfrak{m}}$.
The monotonicity of $\mathfrak{m} \mapsto Q_{\mathfrak{m}}$ then follows by dif\/ferentiation
of~\eqref{ecriture equation definition endpoint Fermi zone dressed momentum}.

\subsection[Large-$Q$ behaviour of $\rho(Q|Q)$]{Large-$\boldsymbol{Q}$ behaviour of $\boldsymbol{\rho(Q|Q)}$}

Proceeding exactly as in the large-$Q$ analysis of $Z(Q|Q)$, namely splitting the integrand of
$R^{(0)}(\lambda,\mu)$ in two parts and computing the contributions coming from the functions $\alpha_-$, we obtain
\begin{gather*}
\rho(Q|Q)-K\big(Q|\gamma/2\big)
=-\int_{\mathbb{R}+{\rm i}\kappa}{}\frac{\text{d} \eta}{4 {\rm i}\pi^2}
\int_{}{}\text{d} \xi \alpha_+(\eta)\cdot \frac{\alpha_+^{-1}(\xi)-1}{\xi-\eta} \int_{-Q}^{Q}\text{d} \mu \cdot K(\mu|\gamma/2)
 e^{{\rm i}(\mu-Q)\eta}
\\
\qquad
- \int_{\mathbb{R}+{\rm i}\kappa}{}
\frac{\text{d} \eta}{4 {\rm i}\pi^2} \int_{}{}\text{d}\xi\alpha_-^{-1}(\eta)\cdot\frac{\alpha_+(\xi)-1}{\xi-\eta} e^{-2 {\rm i}\xi Q}
\int_{-Q}^{Q}\text{d}\mu\cdot K(\mu|\gamma/2)e^{{\rm i}(\mu-Q)\eta}+O\big(e^{-2\epsilon_{\gamma}Q}\big).
\end{gather*}
The $\xi$-integral in the f\/irst line can be explicitly evaluated whereas the integral term in the second line is
already $O\big(e^{-2\kappa Q}\big)$.
Thus, all in all,
\begin{gather*}
\rho(Q|Q)-K\big(Q|\gamma/2\big)=\int_{\mathbb{R}+{\rm i}\kappa}{} \frac{\text{d} \eta}{2\pi}
\left\{\big(\alpha_{+}(\eta)-1\big) \int_{-Q}^{Q} \text{d} \mu \cdot
K(\mu|\gamma/2) e^{{\rm i}(\mu-Q)\eta} \right\}+ O(\delta_{Q})
\\
\qquad
=-\int_{\mathbb{R}}{} \frac{\text{d} \eta}{2\pi} \int_{\mathbb{R}}{}\frac{\text{d}\xi}{2{\rm i}\pi}\left\{\big(\alpha_{+}(\eta)-1\big)\cdot
\mathcal{F}\big[ K(*|\gamma/2)\big](\xi) \cdot\frac{ e^{-{\rm i}\xi Q}- e^{{\rm i}\xi Q-2 {\rm i}\eta Q}}{\xi-\eta}\right\}+O(\delta_{Q})
\\
\qquad
=\int_{\mathbb{R}}{}\big(\alpha_{+}(\xi)-1\big)\cdot \mathcal{F}\big[ K(*|\gamma/2)
\big](\xi) \cdot e^{- {\rm i}\xi Q} \cdot \frac{\text{d} \xi}{2\pi}+ O(\delta_{Q}),
\end{gather*}
where we have set $\delta_{Q}= e^{-2 Q \kappa}+ e^{-2 Q \epsilon_{\gamma}}$

Upon identifying $-K\big(Q|\gamma/2\big)$, using the jump condition for $\alpha_{+}(\xi)$ and remarking that
$\kappa$ is, at most, equal to the imaginary part of the zero of $1+\mathcal{F}[K]$ closest to $\mathbb{R}$,  viz.\
to $\epsilon_{\gamma}$, one is led to the representation
\begin{gather}
\rho(Q|Q)= \int_{\mathbb{R}}{} \frac{\alpha_{-}(\xi) \cdot e^{- {\rm i}\xi Q}}{2\cosh(\gamma\xi/2)} \cdot
\frac{\text{d}\xi}{2\pi} +O\big(e^{-2 Q \epsilon_{\gamma}}\big)
=\frac{e^{-\frac{\pi Q}{\gamma}}}{\gamma} \cdot \alpha(-{\rm i} \pi/\gamma) +O(\delta_{Q}).
\label{ecriture asymptotiques densite en Q}
\end{gather}

Therefore, for
$
2\epsilon_{\gamma}>\frac{\pi}{\gamma}$
viz.\
$\gamma>\frac{\pi}{5}$,
the f\/irst term in the r.h.s.\
of~\eqref{ecriture asymptotiques densite en Q} does correspond to the f\/irst correction.
In the regime $0<\gamma<\pi/5$, one most probably reaches the same conclusion.
However, the analysis is much more technical, so we do not discuss it here.

\section{The dressed energy}
\label{Section dressed energy proofs}

\subsection{Main properties and Fermi rapidity}

The regularity properties of the solution $\varepsilon(\lambda|Q)$ are readily read of\/f from the integral equation
and the smoothness properties of the resolvent $R_{Q}(\lambda,\mu)$.
We shall now establish the bounds on $\varepsilon$.
We start with the bound involving $\widetilde{\varepsilon}$ since it follows readily from the previous considerations.
Solving the integral equation for $\varepsilon$ in terms of the afore-introduced functions leads to the expression
\begin{gather*}
\varepsilon(\lambda|Q)=h Z(\lambda|Q)-4\pi J \sin(\gamma) \rho(\lambda|Q).
\end{gather*}
It then solely remains to apply the (upper or lower, depending on the value of $\gamma$) bounds~\eqref{ecriture bornes
pour Z ga low} and~\eqref{ecriture bornes pour rho ga low} or~\eqref{ecriture bornes pour rho ga high}
and~\eqref{ecriture bornes pour Z ga high} so as to get the bounds involving $\widetilde{\varepsilon}(\lambda)$.
We thus focus on the bounds involving $\varepsilon_0$.
Starting from the representation
\begin{gather*}
\varepsilon(\lambda|Q)=\varepsilon_0(\lambda)-\int_{-Q}^{Q} R_{Q}(\lambda,\mu) \varepsilon_0(\mu)\cdot \text{d} \mu
\end{gather*}
and using that $R_{Q}(\lambda,\mu) >0$ for $0 < \gamma < \pi/2$ and $R_{Q}(\lambda,\mu) < 0$ for $\pi/2 < \gamma < \pi$,
the bounds follow since $-\varepsilon_0(\lambda) > 0$ for any $\lambda \in ]-Q;Q[$ as soon as $Q < Q_0$.

We now turn to the proof of existence and uniqueness of the Fermi rapidity.
In the regime $0<\gamma<\pi/2$, we repeat the change of integration domain trick so as to recast the linear integral
equation satisf\/ied by $\varepsilon$ in the form
\begin{gather*}
\varepsilon(\lambda|Q)-\int_{\mathbb{R} \setminus [-Q;Q]}{}
R(\lambda-\mu)\cdot \varepsilon(\mu|Q) \cdot \text{d} \mu =\varepsilon_{\infty}(\lambda)
\end{gather*}
with
\begin{gather*}
\varepsilon_{\infty}(\lambda)=\lim_{Q \rightarrow +\infty}\varepsilon(\lambda|Q)
=\frac{h\pi}{2(\pi-\gamma)}-\frac{2\pi J\sin(\gamma)}{\gamma\cosh\big(\frac{\pi\lambda}{\gamma}\big)}.
\end{gather*}
Let, $\mathcal{R}(\lambda,\mu)$ be the resolvent kernel to the operator $I-R$ understood as acting on $
\mathcal{C}^{0}_{b}(\mathbb{R}\setminus[-Q;Q])$.
As already argued, it is strictly positive.
Furthermore, manipulations similar to those already discussed lead to
\begin{gather*}
\partial_{\lambda}\varepsilon(\lambda|Q)=\varepsilon(Q|Q) [ \mathcal{R}(\lambda,Q)-\mathcal{R}(\lambda,-Q)]
+(I+\mathcal{R})[\varepsilon^{\prime}_{\infty}](\lambda),
\\
\partial_{Q}\varepsilon(\lambda|Q) =- \varepsilon(Q|Q) [\mathcal{R}(\lambda,Q)+\mathcal{R}(\lambda,-Q)]
\end{gather*}
 viz.\
\begin{gather}
\frac{\text{d}}{\text{d} Q}\varepsilon(Q|Q) =- 2\varepsilon(Q|Q)\mathcal{R}(Q,-Q)+\varepsilon^{\prime}_{\infty}(\lambda)
+\int_{Q}^{+\infty}[\mathcal{R}(\lambda,\nu)-\mathcal{R}(\lambda,-\nu)]\cdot\varepsilon^{\prime}_{\infty}(\nu) \cdot \text{d} \nu,
\label{ecriture derivee totale veps Q Q}
\end{gather}
where we have used that $\varepsilon^{\prime}_{\infty}$ is odd.
Since $R(\lambda-\mu)-R(\lambda+\mu)>0$ for $\lambda,\mu \in [Q;+\infty[$, a~direct inspection of the dif\/ference of
Neumann series (see~\cite[Lemma 2.3]{KozProofexistenceAEYangYangEquation} for more details) shows that
$\big[\mathcal{R}(\lambda,\nu)-\mathcal{R}(\lambda,-\nu)\big]>0$.
As $\varepsilon_{\infty}^{\prime}(\lambda)>0$ for $\lambda\in\mathbb{R}^+$, one gets that all but the f\/irst term in
the  r.h.s.\ of~\eqref{ecriture derivee totale veps Q Q} are strictly positive.
As a~consequence, every zero of $Q\mapsto \varepsilon\big(Q|Q\big)$ belongs to an open set on which the function is increasing.
Since $Q \mapsto \varepsilon(Q|Q) $ is a~continuous function, it has thus at most one zero.
Further, the bounds $\widetilde{\varepsilon}(Q)>\varepsilon(Q|Q)>\varepsilon_0(Q)$ ensure that
$\varepsilon\big(Q_0|Q_0\big)> 0$ and $\varepsilon\big(\widetilde{Q}|\widetilde{Q}\big) < 0$.
Thus, by continuity, there exists $Q_F \in ] \widetilde{Q}; Q_0 [$ such that $\varepsilon(Q_F|Q_F) =0$.

It remains to treat the case $\pi/2 < \gamma < \pi$.
Similar calculations lead to
\begin{gather*}
\frac{\text{d}}{\text{d} Q}\varepsilon(Q|Q)=- 2\varepsilon(Q|Q) R_{Q}(Q, -Q)\big]\cdot \varepsilon^{\prime}_{0}(\nu)
-\int_{0}^{Q}[R_Q(\lambda,\nu)-R_Q(\lambda,-\nu)]\cdot \varepsilon^{\prime}_{0}(\nu)\cdot \text{d} \nu.
\end{gather*}
Since, as follows from Lemma~\ref{Lemme bounds up and low resolvent}, $-\big[ R_Q(\lambda,\nu)-R_Q(\lambda,-\nu)\big]>0$,
one concludes as before regarding to uniqueness of $Q_F$.
Its existence follows from the bounds $\varepsilon_0(Q)>\varepsilon(Q|Q)> \widetilde{\varepsilon}(Q)$.

It remains to justify the smoothness of the map $h \mapsto Q_F$.
The latter follows from the implicit function theorem, while the expression for $ \partial_{h} Q_F(h)$ follows from
straightforward calculations.

\subsection{Positivity of the real part on the other bank}

We close the section devoted to the dressed energy by establishing a~strictly positive, magnetic f\/ield dependent,
lower bound on $\Re(\varepsilon_{+}(\lambda- {\rm i}\gamma))$, $\lambda \in \mathbb{R}$.
In order to prove this result, we f\/irst need to establish a~technical lemma.

\begin{lemma}
\label{Lemme proprietes fct G}
The function ${\cal G}$ defined as the Fourier transform
\begin{gather}
\label{fintg}
{\cal G} (\lambda)=\int_{\mathbb{R}}{} \frac{\sinh[(\pi/ 2-2\gamma)k] e^{-{\rm i}k\lambda}}{\cosh\big(\gamma k/ 2\big)\sinh[(\pi/ 2-\gamma/ 2)k]}\cdot
\frac{\text{\rm d}k}{4\pi}
\end{gather}
has the following properties:
\begin{enumerate}[$(i)$]
\item
For $0 < \gamma < \frac{\pi}{4}$ the function $\cal G$ is real, even and positive on the real axis and monotonically
decreasing with vanishing asymptotics for $\lambda > 0$.
\item
For $\frac{\pi}{4} < \gamma < \frac{\pi}{2}$ the function $\cal G$ is real, even and negative on the real axis and has
vanishing asymptotics for $\lambda > 0$.
\end{enumerate}
\end{lemma}

\begin{proof}
In order to prove (i) and (ii) we def\/ine
\begin{gather*}
\gamma''=\gamma''(\gamma)=\frac{3 \gamma/2}{1-\gamma/\pi}.
\end{gather*}
This function of $\gamma$ is monotonically increasing for $0 < \gamma < \pi$, and
\begin{gather*}
\gamma''(\pi/4)=\pi/2,
\qquad
\gamma''(2\pi/5)=\pi.
\end{gather*}
The Fourier transformation formulae~\eqref{Ecriture Fourier K et 1+K} only hold as long as $\gamma < \pi$.
Thus, we obtain a~representation as a~convolution, similar to~\eqref{convreff}, only for $0 < \gamma < 2\pi/5$.
For $0 < \gamma < \pi/4$, we have $0 < \gamma'' < \pi/2$ and
\begin{gather}
\label{convrefg}
{\cal G} (\lambda) = \frac{\pi}{2\gamma (\pi-\gamma)} \int_{\mathbb{R}}{}
\frac{K\big(\frac{\mu}{1-\gamma/ \pi}\big|\gamma''\big)}{\cosh [\pi(\lambda-\mu)/\gamma]} \cdot \text{d} \mu > 0.
\end{gather}
For $\pi/4 < \gamma < 2\pi/5$ we have the same representation~\eqref{convrefg} but with $\pi/2 < \gamma'' < \pi$.
In this range $K(\lambda|\gamma'')$ is negative, hence ${\cal G} (\lambda) < 0$.
Using the same reasoning as above we can also say that $\cal G$ is even, asymptotically vanishing and, for $\lambda>0$,
monotonically decreasing if $0 < \gamma < \pi/4$ and monotonically increasing if $\pi/4 < \gamma < 2\pi/5$.

If $\gamma > 2\pi/5$ a~representation like~\eqref{convrefg} does not exist anymore, since then the inverse Fourier
transform of $ \sinh\big[ (\pi/2-2\gamma)k\big] / \sinh\big[ (\pi/2 -\gamma/2)k\big] $ does not exist.
Still, using the elementary identity
\begin{gather*}
\frac{\sinh\big[\big(\frac{\pi}{2}-2\gamma\big)k\big]}{2\cosh\big[\frac{\gamma k}{2}\big]\sinh\big[\big(\frac{\pi}{2}-\frac{\gamma}{2}\big)k\big]}
=\frac{\sinh\big[\big(\frac{\pi}{2}-\frac{3 \gamma}{2}\big)k\big]}{\sinh\big[\big(\frac{\pi}{2}-\frac{\gamma}{2}\big)k\big]}
-\frac{\sinh\big[\big(\frac{\pi}{2}-\gamma\big)k\big]}{2\cosh\big(\frac{\gamma k} 2\big) \sinh\big[\big(\frac{\pi}{2}-\frac \gamma 2\big)k\big]}
\end{gather*}
we f\/ind that
\begin{gather}
\label{repg2}
{\cal G} (\lambda)=\Big(1-\frac{\gamma}{\pi}\Big)^{-1} \cdot K\left(\frac{\lambda \cdot \pi}{\pi-\gamma}\bigg|\gamma'''\right)-R (\lambda),
\end{gather}
where $\gamma''' = 2\gamma'$ and $\gamma'$ has been def\/ined in~\eqref{convreff}.
Again $\gamma'''$ is a~monotonically increasing function of $\gamma$ for $0 < \gamma < \pi$, but now
\begin{gather*}
\gamma''' (\pi/3)=\pi/2,
\qquad
\gamma''' (\pi/2) = \pi.
\end{gather*}
It follows from~\eqref{repg2} that ${\cal G} (\lambda) < 0$ for $\pi/3 < \gamma < \pi/2$.
Since we had already shown that~${\cal G}$ is negative for $\pi/4 < \gamma < 2\pi/5$ and since $\pi/3 < 2\pi/5$, we
conclude that ${\cal G} (\lambda) < 0$ for $\pi/4 < \gamma < \pi/2$.
\end{proof}

We are f\/inally in position to prove Proposition~\ref{Proposition Dressed energy positivity on other bank}:

\begin{proof}
It follows from~\eqref{Definition equation dressed charge} that $\varepsilon$ is meromorphic in the strip $|\Im
(\lambda) | < \gamma$ with simple poles at $\pm {\rm i}\gamma/2$ and that
\begin{gather*}
\text{Res}\big(\varepsilon (\lambda|Q_F) \cdot \text{d} \lambda, \lambda = \mp {\rm i}\gamma/2\big)
=\mp 2 {\rm i}J \sin (\gamma).
\end{gather*}
Then, by deforming the contour in~\eqref{Definition equation dressed charge} we obtain, for $\lambda \in {\mathbb R}$,
\begin{gather*}
\varepsilon_+ (\lambda-{\rm i}\gamma|Q_F)=\varepsilon_0 (\lambda-{\rm i}\gamma)+4 \pi J \sin (\gamma)
K(\lambda-{\rm i}\gamma/2)
\\
\qquad
+\int_{{\mathbb R} \setminus [-Q_F;Q_F]}{}K(\lambda-\mu-{\rm i}\gamma+{\rm i}0)\cdot\varepsilon
(\mu|Q_F)\cdot\text{d}\mu-\int_{\mathbb{R}}{}K(\lambda-\mu)\varepsilon_+(\mu-{\rm i}\gamma|Q_F)\cdot\text{d}\mu.
\end{gather*}
Setting
\begin{gather*}
\omega(\lambda)=\Re[\varepsilon_+ (\lambda-{\rm i}\gamma|Q_F)]
\end{gather*}
and using
\begin{gather*}
K (\lambda-{\rm i}\gamma+{\rm i}0) = \frac{\delta (\lambda)}{2}-\frac{1}{2\pi {\rm i}}
\text{p.v.} \coth(\lambda)+\frac{1}{2\pi {\rm i}} \coth(\lambda-2 {\rm i}\gamma),
\\
\varepsilon_0 (\lambda-{\rm i}\gamma)+4 \pi J \sin (\gamma) K(\lambda-{\rm i}\gamma/2)=h+4 \pi J \sin (\gamma)
K(\lambda|\gamma/2)
\end{gather*}
for $\lambda \in {\mathbb R}$, we obtain
\begin{gather}
\omega (\lambda) = h+4 \pi J \sin (\gamma) K(\lambda|\gamma/2) +\frac{\varepsilon_{c}(\lambda)}{2}
\nonumber
\\
\phantom{\omega (\lambda)=}
{}+\frac{1}{2} \int_{\mathbb{R}}{} K(\lambda-\mu|2\gamma) \cdot \varepsilon_c (\mu) \cdot \text{d} \mu
-\int_{\mathbb{R}}{} K(\lambda-\mu) \omega (\mu) \cdot \text{d} \mu,
\label{omint}
\end{gather}
which is an equation for real functions on the real line.
Above, we agree upon
\begin{gather*}
\varepsilon_c (\lambda) = \varepsilon (\lambda|Q_F) \cdot \Theta (|\lambda|-Q_F),
\end{gather*}
where $\Theta$ is the Heaviside step function.
Equation~\eqref{omint} can be solved by means of Fourier transformation.
Using~\eqref{Ecriture Fourier K et 1+K} and
\begin{gather*}
\frac{\mathcal{F}[K(*|\gamma/2)](k)}{1+\mathcal{F}[K](k)}=\frac{1}{2\cosh(\gamma k/2)}
\end{gather*}
we obtain
\begin{gather}
\omega (\lambda)=\frac{h \pi}{2 (\pi-\gamma)}+\frac{2\pi J \sin (\gamma)/\gamma}{\cosh\big(\frac{\pi \lambda}\gamma\big)} +\frac{\varepsilon_c (\lambda)}{2}
\nonumber
\\
\phantom{\omega (\lambda)=}{}
+\frac{1}{2} \int_{{\mathbb R} \setminus [-Q_F;Q_F]}{}\{{\cal G} (\lambda-\mu)-R (\lambda-\mu)\} \cdot \varepsilon(\mu|Q_F) \cdot\text{d} \mu.
\label{om}
\end{gather}
Here, we recall that $R$ and $\cal G$ are def\/ined by the Fourier integrals~\eqref{fintr} and~\eqref{fintg}.

By construction, the function $\varepsilon$ is positive on ${\mathbb R} \setminus [-Q_F;Q_F]$, where it is also bounded
from above by $h$.
We need to f\/ind a~lower bound for the integral on the right hand side of~\eqref{om}.
For this purpose we distinguish two cases.
\begin{itemize}\itemsep=0pt
\item
$0< \gamma \leq \pi/4$.
Then, according to Lemmas~\ref{Lemme: propr R} and~\ref{Lemme proprietes fct G},
\begin{gather}
\frac{1}{2}\int_{{\mathbb R} \setminus [-Q_F;Q_F]}{}\{{\cal G}(\lambda-\mu)-R(\lambda-\mu)\} \varepsilon(\mu|Q_F) \cdot \text{d} \mu
\nonumber
\\
\qquad
\geq-\frac{1}{2}\int_{{\mathbb R} \setminus [-Q_F;Q_F]}{}R (\lambda-\mu) \cdot\varepsilon(\mu) \cdot \text{d} \mu
>-\frac{h}{2} \int_{\mathbb{R}}{} R (\lambda) \cdot \text{d} \lambda=-\frac{h}{2} \frac{\frac{1}{2}-\frac{\gamma}{\pi}}{1-\frac{\gamma}{\pi}}. \label{est1}
\end{gather}

\item
$\pi/4 \leq \gamma < \pi/2$.
Then, according to Lemmas~\ref{Lemme: propr R} and~\ref{Lemme proprietes fct G},
\begin{gather}
\frac{1}{2}\int_{{\mathbb R} \setminus [-Q_F;Q_F]}{}\Big\{{\cal G} (\lambda-\mu)-R (\lambda-\mu)\Big\} \cdot \varepsilon(\mu|Q_F) \cdot \text{d} \mu
\nonumber
\\
\qquad
>\frac{h}{2} \int_{\mathbb{R}}{}\{{\cal G} (\lambda)-R (\lambda)\} \cdot \text{d} \lambda=- \frac{h}{2} \frac{\gamma/\pi}{1-\frac{\gamma}{\pi}}.
\label{est2}
\end{gather}
\end{itemize}

\noindent Using~\eqref{est1},~\eqref{est2} in~\eqref{om} we obtain
\begin{gather*}
\omega (\lambda)>\frac{2\pi J \sin (\gamma)/\gamma}{\cosh (\pi \lambda/\gamma)}+\frac{\varepsilon_c (\lambda)}{2}
+\frac{h}{2}
\begin{cases}
\dfrac{\frac{1}{2}+\frac{\gamma}{\pi}}{1-\frac{\gamma}{\pi}} & \text{for}
\quad
0 < \gamma \leq \pi/4,
\\
1 & \text{for}
\quad
\pi/4 \leq \gamma < \pi/2,
\end{cases}
\end{gather*}
which implies~\eqref{Ecriture lower bound on other bank}.
\end{proof}

\subsection[Dependence of $Q_F(h)$ on $h$ for small $h$]{Dependence of $\boldsymbol{Q_F(h)}$ on $\boldsymbol{h}$ for small $\boldsymbol{h}$}

We have established that $h \mapsto Q_F(h)$ is a~strictly decreasing function of $h$.
Thence, since $\varepsilon\big(\lambda|+\infty\big)<0$ on $\mathbb{R}$, it follows that $\lim\limits_{h \rightarrow
0^+}Q_F(h) =+\infty$.
Hence, for $h$ small enough,
\begin{gather*}
\varepsilon\big(Q_F(h)|Q_F(h)\big)=h \alpha_{+}(0)-4\pi J \frac{\sin \gamma}{\gamma} e^{- \pi Q_F(h)/\gamma}\alpha_-(-{\rm i}\pi/\gamma)
+O\big(e^{-2\varepsilon_{\gamma} Q_F(h)}\big).
\end{gather*}
Upon explicating the values of $\alpha$ one gets the claimed form of the small-$h$ asymptotics.

\section{Conclusion}

In the present paper, we have proved several properties of solutions to linear integral equations arising in the
description of the ground state of the XXZ spin-1/2 chain in the thermodynamic limit.
Although we have focused our analysis on this specif\/ic model, we do trust that the method is more general and can be
applied to other linear integral equations arising in the context of the thermodynamic limit of more complex quantum
integrable models.
In particular, with minor modif\/ications, the techniques should work for models based on a~$\mathcal{Y}(\mathfrak{g})$
or~$U_{q}(\mathfrak{g})$ symmetry, with~$\mathfrak{g}$ a~Lie algebra of rank higher than 1.

\subsection*{Acknowledgements}

The authors would like to thank J.~Suzuki for helpful discussions.
MD and FG acknowledge f\/inancial support by the Volkswagen foundation and by the DFG under grant \# Go 825/7-1.
KKK is supported by the CNRS.
His work has been partly f\/inanced by the Burgundy region PARI 2013 FABER grant `Structures et asymptotiques d'int\'egrales multiples'.

\pdfbookmark[1]{References}{ref}
\LastPageEnding


\begin{thebibliography}{99}
\footnotesize\itemsep=0pt

\bibitem{BLZ97}
Bazhanov V.V., Lukyanov S.L., Zamolodchikov A.B., Integrable structure of
  conformal f\/ield theory. {II}.~{${\rm Q}$}-operator and {DDV} equation,
  \href{http://dx.doi.org/10.1007/s002200050240}{\textit{Comm. Math. Phys.}} \textbf{190} (1997), 247--278,
  \href{http://arxiv.org/abs/hep-th/9604044}{hep-th/9604044}.

\bibitem{BetheSolutionToXXX}
Bethe H., Zur Theorie der Metalle: Eigenwerte und Eigenfunktionen der linearen
  Atomkette, \href{http://dx.doi.org/10.1007/BF01341708}{\textit{Z.~Phys.}} \textbf{71} (1931), 205--226.

\bibitem{BJMS10}
Boos H., Jimbo M., Miwa T., Smirnov F., Hidden {G}rassmann structure in the
  {XXZ} model {IV}: {CFT} limit, \href{http://dx.doi.org/10.1007/s00220-010-1051-6}{\textit{Comm. Math. Phys.}} \textbf{299}
  (2010), 825--866, \href{http://arxiv.org/abs/0911.3731}{arXiv:0911.3731}.

\bibitem{DeVegaWoynarowichFiniteSizeCorrections6VertexNLIEmethod}
de~Vega H.J., Woynarovich F., Method for calculating f\/inite size corrections in
  {B}ethe ansatz systems: {H}eisenberg chain and six-vertex model,
  \href{http://dx.doi.org/10.1016/0550-3213(85)90271-8}{\textit{Nuclear Phys.~B}} \textbf{251} (1985), 439--456.

\bibitem{DescloizeauxGaudinExcitationsXXZ+Gap}
des Cloizeaux J., Gaudin M., Anisotropic linear magnetic chain,
  \href{http://dx.doi.org/10.1063/1.1705048}{\textit{J.~Math. Phys.}} \textbf{7} (1966), 1384--1400.

\bibitem{DescloizeauxPearsonExcitationsXXX}
des Cloizeaux J., Pearson J.J., Spin-wave spectrum of the antiferromagnetic
  linear chain, \href{http://dx.doi.org/10.1103/PhysRev.128.2131}{\textit{Phys. Rev.}} \textbf{128} (1962), 2131--2135.

\bibitem{DesVeg95}
Destri C., de~Vega H.J., Unif\/ied approach to thermodynamic {B}ethe ansatz and
  f\/inite size corrections for lattice models and f\/ield theories,
  \href{http://dx.doi.org/10.1016/0550-3213(94)00547-R}{\textit{Nuclear Phys.~B}} \textbf{438} (1995), 413--454,
  \href{http://arxiv.org/abs/hep-th/9407117}{hep-th/9407117}.

\bibitem{DorlasSamsonovThermoLim6VertexAndConvergceToDensityInSomeCases6VrtX}
Dorlas T.C., Samsonov M., On the thermodynamic limit of the 6-vertex model,
  \href{http://arxiv.org/abs/0903.2657}{arXiv:0903.2657}.


\bibitem{KozDugaveGohmannThermaxFormFactorsXXZ}
Dugave M., G\"{o}hmann F., Kozlowski K.K., Thermal form factors of the XXZ
  chain and the large-distance asymptotics of its temperature dependent
  correlation functions, \href{http://dx.doi.org/10.1088/1742-5468/2013/07/P07010}{\textit{J.~Stat. Mech. Theory Exp.}} \textbf{2013}
  (2013), P06002, 52 pages, \href{http://arxiv.org/abs/1305.0118}{arXiv:1305.0118}.

\bibitem{DGK14app}
Dugave M., G\"{o}hmann F., Kozlowski K.K., Low-temperature large-distance
  asymptotics of the transversal two-point functions of the {XXZ} chain,
  \href{http://arxiv.org/abs/1401.4132}{arXiv:1401.4132}.

\bibitem{EsslerFrahmGohmanKlumperKorepinOneDimensionalHubbardModel}
Essler F.H.L., Frahm H., G\"{o}hmann F., Kl\"{u}mper A., Korepin V.E., The
  one-dimensional Hubbard model, Cambridge University Press, Cambridge, 2005.

\bibitem{GaudinTBAXXZMassiveInfiniteSetNLIE}
Gaudin M., Thermodynamics of a Heisenberg--Ising ring for $\Delta \geq 1$,
  \href{http://dx.doi.org/10.1103/PhysRevLett.26.1301}{\textit{Phys. Rev. Lett.}} \textbf{26} (1971), 1301--1304.

\bibitem{GriffithsXXZFirstLinIntEqnFiniteMagField}
Grif\/f\/iths R.B., Magnetization curve at zero temperature for the
  antiferromagnetic Heisenberg linear chain, \href{http://dx.doi.org/10.1103/PhysRev.133.A768}{\textit{Phys. Rev.}} \textbf{133}
  (1964), 768--775.

\bibitem{HultenGSandEnergyForXXX}
Hulth\'en L., \"{U}ber das Austauschproblem eines Kristalles, \textit{Arkiv
  Mat. Astron. Fys.~A} \textbf{26} (1938), 1--106.

\bibitem{KozKitMailSlaTerXXZsgZsgZAsymptotics}
Kitanine N., Kozlowski K.K., Maillet J.M., Slavnov N.A., Terras V., Algebraic
  {B}ethe ansatz approach to the asymptotic behavior of correlation functions,
  \href{http://dx.doi.org/10.1088/1742-5468/2009/04/P04003}{\textit{J.~Stat. Mech. Theory Exp.}} \textbf{2009} (2009), P04003, 66~pages,
  \href{http://arxiv.org/abs/0808.0227}{arXiv:0808.0227}.

\bibitem{KozKitMailSlaTerEffectiveFormFactorsForXXZ}
Kitanine N., Kozlowski K.K., Maillet J.M., Slavnov N.A., Terras V., On the
  thermodynamic limit of form factors in the massless {XXZ} {H}eisenberg
  chain, \href{http://dx.doi.org/10.1063/1.3136683}{\textit{J.~Math. Phys.}} \textbf{50} (2009), 095209, 24~pages,
  \href{http://arxiv.org/abs/0903.2916}{arXiv:0903.2916}.

\bibitem{KozKitMailSlaTerRHPapproachtoSuperSineKernel}
Kitanine N., Kozlowski K.K., Maillet J.M., Slavnov N.A., Terras V.,
  Riemann--{H}ilbert approach to a generalised sine kernel and applications,
  \href{http://dx.doi.org/10.1007/s00220-009-0878-1}{\textit{Comm. Math. Phys.}} \textbf{291} (2009), 691--761, \href{http://arxiv.org/abs/0805.4586}{arXiv:0805.4586}.

\bibitem{KlumperNLIEfromQTMDescrThermoXYZOneUnknownFcton}
Kl\"{u}mper A., Thermodynamics of the anisotropic spin-$1/2$ Heisenberg chain
  and related quantum chains, \href{http://dx.doi.org/10.1007/BF01316831}{\textit{Z.~Phys.~B}} \textbf{91} (1993), 507--519,
  \href{http://arxiv.org/abs/cond-mat/9306019}{cond-mat/9306019}.

\bibitem{KlumperBatchelorNLIEApproachFiniteSizeCorSpin1XXZIntroMethod}
Kl{\"u}mper A., Batchelor M.T., An analytic treatment of f\/inite-size
  corrections in the spin-{$1$} antiferromagnetic {XXZ} chain,
  \href{http://dx.doi.org/10.1088/0305-4470/23/5/002}{\textit{J.~Phys.~A: Math. Gen.}} \textbf{23} (1990), L189--L195.

\bibitem{KlumperWehnerZittartzConformalSpectrumofXXZCritExp6Vertex}
Kl{\"u}mper A., Wehner T., Zittartz J., Conformal spectrum of the six-vertex
  model, \href{http://dx.doi.org/10.1088/0305-4470/26/12/021}{\textit{J.~Phys.~A: Math. Gen.}} \textbf{26} (1993), 2815--2827.

\bibitem{BogoliubiovIzerginKorepinBookCorrFctAndABA}
Korepin V.E., Bogoliubov N.M., Izergin A.G., Quantum inverse scattering method
  and correlation functions, \href{http://dx.doi.org/10.1017/CBO9780511628832}{\textit{Cambridge Monographs on Mathematical Physics}},
  Cambridge University Press, Cambridge, 1993.

\bibitem{KozProofexistenceAEYangYangEquation}
Kozlowski K.K., Low-{$T$} asymptotic expansion of the solution to the
  {Y}ang--{Y}ang equation, \href{http://dx.doi.org/10.1007/s11005-013-0654-1}{\textit{Lett. Math. Phys.}} \textbf{104} (2014),
  55--74, \href{http://arxiv.org/abs/1112.6199}{arXiv:1112.6199}.

\bibitem{KulishReshetikhinNestedBAFirstIntroduction}
Kulish P.P., Reshetikhin N.Yu., Generalized {H}eisenberg ferromagnet and the
  {G}ross--{N}eveu model, \textit{Soviet Phys. JETP} \textbf{80} (1981),
  214--228.

\bibitem{LiebWuFirstDerivationOfLIE4Hubbard1D}
Lieb E.H., Wu F.Y., Absence of Mott transition in an exact solution of the
  short-range, one-band model in one dimension, \href{http://dx.doi.org/10.1103/PhysRevLett.20.1445}{\textit{Phys. Rev. Lett.}}
  \textbf{20} (1968), 1445--1448, {E}rratum,
  \href{http://dx.doi.org/10.1103/PhysRevLett.21.192.2}{\textit{Phys. Rev.
  Lett.}} \textbf{21} (1968), 192.

\bibitem{OrbachXXZCBASolution}
Orbach R., Linear antiferromagnetic chain with anisotropic coupling,
  \href{http://dx.doi.org/10.1103/PhysRev.112.309}{\textit{Phys. Rev.}} \textbf{112} (1958), 309--316.

\bibitem{TakahashiTBAforXXZFiniteTinfiniteNbrNLIE}
Takahashi M., One-dimensional Heisenberg model at f\/inite temperature,
  \href{http://dx.doi.org/10.1143/PTP.46.401}{\textit{Progr. Theoret. Phys.}} \textbf{46} (1971), 401--415.

\bibitem{WalkerFirstExplicitSolToLinIntEqnMassiceXXZ}
Walker L.R., Antiferromagnetic linear chain, \href{http://dx.doi.org/10.1103/PhysRev.116.1089}{\textit{Phys. Rev.}} \textbf{116}
  (1959), 1089--1090.

\bibitem{Yang-YangXXZStructureofGS}
Yang C.N., Yang C.P., One-dimensional chain of anisotropic spin-spin
  interactions. II.~Properties of the ground-state energy per lattice site for
  an inf\/inite system, \href{http://dx.doi.org/10.1103/PhysRev.150.327}{\textit{Phys. Rev.}} \textbf{150} (1966), 327--339.

\bibitem{Yang-YangNLSEThermodynamics}
Yang C.N., Yang C.P., Thermodynamics of a one-dimensional system of bosons with
  repulsive delta-function interaction, \href{http://dx.doi.org/10.1063/1.1664947}{\textit{J.~Math. Phys.}} \textbf{10}
  (1969), 1115--1122.

\bibitem{Zamolodchikov90}
Zamolodchikov Al.B., Thermodynamic {B}ethe ansatz in relativistic models:
  scaling {$3$}-state {P}otts and {L}ee--{Y}ang models, \href{http://dx.doi.org/10.1016/0550-3213(90)90333-9}{\textit{Nuclear
  Phys.~B}} \textbf{342} (1990), 695--720.

\end{thebibliography}
\end{document}